\documentclass[pra,twocolumn,showpacs]{revtex4-2}
\usepackage{amsmath,amscd,amssymb,color}
\usepackage{graphicx,amsfonts,dsfont}
\usepackage{epstopdf}
\usepackage{hyperref}
\usepackage{enumerate,bbold}
\usepackage{array}
\usepackage{braket}

\usepackage{ulem}

\begin{document}
\title{Protection of noisy multipartite entangled states of
superconducting qubits via universally robust dynamical decoupling schemes}
\author{Akanksha Gautam}
\email{akankshagautam@iisermohali.ac.in}
\affiliation{Department of Physical Sciences, Indian
Institute of Science Education \& 
Research Mohali, Sector 81 SAS Nagar, 
Manauli PO 140306 Punjab India.}
\author{Arvind}
\email{arvind@iisermohali.ac.in}
\affiliation{Department of Physical Sciences, Indian
Institute of Science Education \& 
Research Mohali, Sector 81 SAS Nagar, 
Manauli PO 140306 Punjab India.}
\author{Kavita Dorai}
\email{kavita@iisermohali.ac.in}
\affiliation{Department of Physical Sciences, Indian
Institute of Science Education \& 
Research Mohali, Sector 81 SAS Nagar, 
Manauli PO 140306 Punjab India.}

\begin{abstract}
We demonstrate the
efficacy of the 
universally robust dynamical decoupling (URDD) sequence 
to preserve multipartite maximally entangled 
quantum states on a cloud
based quantum computer via the IBM platform. 
URDD is a technique that can
compensate for experimental errors and simultaneously 
protect the state against 
environmental noise. To further improve the performance of the URDD sequence,
phase randomization (PR) as well as correlated phase randomization (CPR)
techniques are added to the basic 
URDD sequence. The performance of the URDD sequence is
quantified 
by measuring the entanglement in several noisy entangled states
(two-qubit triplet state, three-qubit GHZ state, four-qubit GHZ state 
and four-qubit
cluster state) at several  time points.  Our
experimental results demonstrate that the URDD sequence is successfully able to
protect noisy multipartite entangled states and its performance is 
substantially
improved by adding the phase randomization and correlated phase randomization
sequences.
\end{abstract}
\maketitle 
\section{Introduction}
\label{intro}
Quantum entanglement is a type of quantum
correlation~\citep{nielsen-2010,schro-35} which has wide applications in
quantum computing and quantum information processing such as quantum secret
sharing~\citep{Hillery-99,theo-2000}, quantum cryptography
\citep{Bennett-tcs-14}, quantum teleportation~\citep{Bennett-93} and quantum
dense coding~\citep{Bennett-92}. However, as the number of qubits increases,
control of the entire system becomes difficult due to the presence of
environmental noise which adversely affects the entanglement present in
multipartite systems~\citep{Yu-prl-2004,Briegel-prl-2004}.  Decoherence due to
system-environment interaction eventually leads to decay of quantumm coherence
and is a major hurdle in building actual quantum computers.  Various strategies
have been proposed to mitigate the effect of decoherence, including quantum
error correction \citep{Shor-pra-1995,steane-prl-96,knill-pra-1997},
decoherence free subspace \citep{Duan-prl-97,Lidar-prl-98} and dynamical
decoupling
(DD)\citep{viola-prl-1999,viola1-prl-1999,Lidar-prl-2005,viola-prl-2005,Lidar-prl-2007,yang-prl-2008,uhrig-prl-2009,Pryadko-pra-2009}.
DD sequences have proved to be successful in protecting the quantum state
without requiring any prior knowledge of the
environment\citep{Du-nat-2009,Wang-prl-2011,Ryan-prl-2010}.

DD is a open-loop control scheme that involves the application of tailored $\pi$
pulses on the system that average out the system-environment
interactions. 
Originally the DD sequence was introduced in the case of nuclear magnetic
resonance (NMR)\citep{Hahn-pr-1950}, from where
this sequence evolved to generate complicated sequences to protect
the quantum state with high precision such as XY4, concatenated DD (CDD), Knill
DD (KDD) and Uhrig DD (UDD) sequences
\citep{Uhrig1-prl-2007,Yang1-prl-2008,Biercuk-pra-2009,Roy-pra-2011,Heinze-prl-2019,Lidar-prl-2007,witzel-prb-2007,Gonzalo-pra-2012}.
The XY4 sequence was proposed to compensate for the errors accumulated from
the repeated application of inverse pulses but it has limited error
compensation. The KDD sequence is a robust sequence against pulse errors derived
from the composite pulses but requires a higher duty cycle for better effect
\citep{Souza-prl-2011}. The CDD sequence is constructed from the recursive pulse
sequence that eliminate decoherence to an arbitrary order but the number 
of pulses
increases exponentially. Thus, this area of research still needs to be explored
to design better DD sequences with limited experimental resources
\citep{Gonzalo-pra-2012}. One such DD sequence called the universally robust
dynamical decoupling (URDD) was recently proposed that can compensate
higher order pulse errors while the number of pulses increases only linearly
and works well for any initial condition \citep{Genov-prl-2017}. Its superior
performance has been verified experimentally in storing coherent optical data
in a Pr:YSO crystal. Further, the protection of single-qubit states has been
shown experimentally using URDD and other previously known DD sequence on
the Rigetti computing platform~\citep{Souza-qip-2021}.

A method was proposed to further improve the robustness of DD sequences, called
the phase randomization technique, in which a random global phase is added to
all $\pi$ pulses of the DD sequence, with this changing each time the unit of
the DD sequence is repeated~\citep{Zhen-prl-2019}.  Another method called the
correlated phase randomization (CPR) technique was proposed which is a variant
of the previously proposed phase randomization (PR) technique, wherein the
fully random phases are replaced with correlated random
phases~\citep{Zhenyu-symm-2020}.  These techniques were originally proposed for
applications in quantum sensing for correct signal identification in
nitrogen-vacancy centers.

Much recent research efforts are focused on increasing the size of quantum
systems, which however are more prone to noise and are hence termed noisy
intermediate-scale quantum (NISQ) computers.  The IBM Quantum Experience is
freely providing such a system based on superconducting qubits to the public
and several researchers have been successful in implementing real quantum
experiments\citep{Harper-prl-2019,sisodia-pla-2017,Lee-sp-2019,Rosi-nqi-2020,Pokharel-prl-2018}.
Therefore, implementing protection schemes on this noisy quantum computer is a
good test of the efficiency of DD sequences in protecting multipartite
entangled states.

In this work, we have experimentally demonstrated the preservation of maximally
entangled multipartite states (two-qubit triplet state, three-qubit GHZ state
and four-qubit GHZ as well as cluster state) generated using superconducting
qubits on the cloud-based IBM quantum experience. These maximally entangled
quantum states are preserved by the application of URDD sequences on the
five-qubit $ibmq_{-}manila$ quantum processor. The  performance of the URDD
sequence is further improved by adding PR as well as CPR sequences. To quantify
the performance of URDD sequence and its variant on adding PR and CPR sequences
in preserving entangled states, we measured the entanglement witness of each
maximally entangled state directly by measuring only a single qubit to verify
the preservation of entanglement on protecting the state.  Our experimental
results show that the URDD sequence is able to successfully preserve the
entanglement of multipartite state. The extent to which the state is preserved
by URDD sequence is further improved by adding phase and correlated phase
randomization techniques.

\begin{figure}[t]
\includegraphics[angle=0,scale=1]{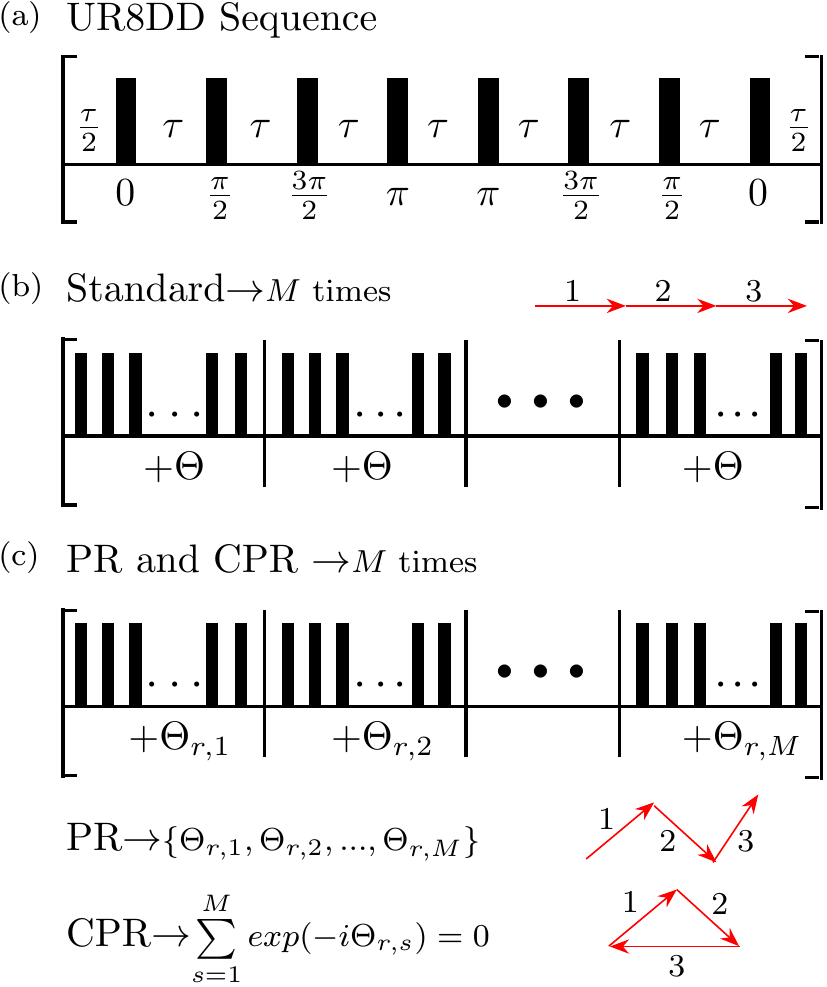}
\caption{(a) A basic unit of a universally robust DD sequence of order $8$
(UR8DD) where the filled black rectangles  
represent $\pi$ pulses. The phase of each pulse is written
below each rectangle and $\tau$ denotes the delay between pulses. (b)
The standard DD protocol where each UR8DD unit is repeated $M$ times,
with a common global
phase $\Theta$ added to each unit of the UR8DD sequence
(example is shown in red). (c) The randomization protocol (PR),  with
a random global phase $\Theta_{r,s}$ being added to each unit of 
the UR8DD sequence; 
in the correlated randomization protocol (CPR), a
constraint is imposed on the random phase (examples of
PR and CPR are shown
in red). } 
\label{urdd8} 
\end{figure} 

This paper is organized as follows: Sec.\ref{urdd} describes the dynamical
decoupling sequences used to protect the entangled states.
Sec.\ref{stu} contains details of the URDD sequence 
and Sec.\ref{prt}  
describes the phase randomization technique, while Sec.\ref{cprt} describes the
correlated phase randomization technique. Sec.\ref{ed} contains the
experimental details of the implementation of the 
URDD sequence on the IBM quantum
processor and Sec.\ref{ew} describes the results of
directly measuring the entanglement witness of the 
two-qubit (triplet state), three-qubit (GHZ state) 
and four-qubit (GHZ and cluster state) maximally entangled
states, after applying the DD protection sequences. Sec.\ref{con}
presents some concluding remarks.

\section{Universally Robust Dynamical Decoupling}
\label{urdd}
Dynamical decoupling (DD) is a technique that consists of sequences of periodic
$\pi$ pulses applied on the quantum system, where each $\pi$ pulse is separated
by free evolution delays of time $\tau$. The phases of $\pi$ pulses and delays
play an important role in the performance of DD sequence where appropriate
choice of phases can make the DD sequence robust against the pulse errors. It
is well known that pulse imperfections are a major factor that limit the
performance of DD sequences. To compensate these pulse errors, the universally
robust DD (URDD) sequence has been recently designed which shows a superior
performance.

\begin{figure*}[t]
\includegraphics[angle=0,scale=1]{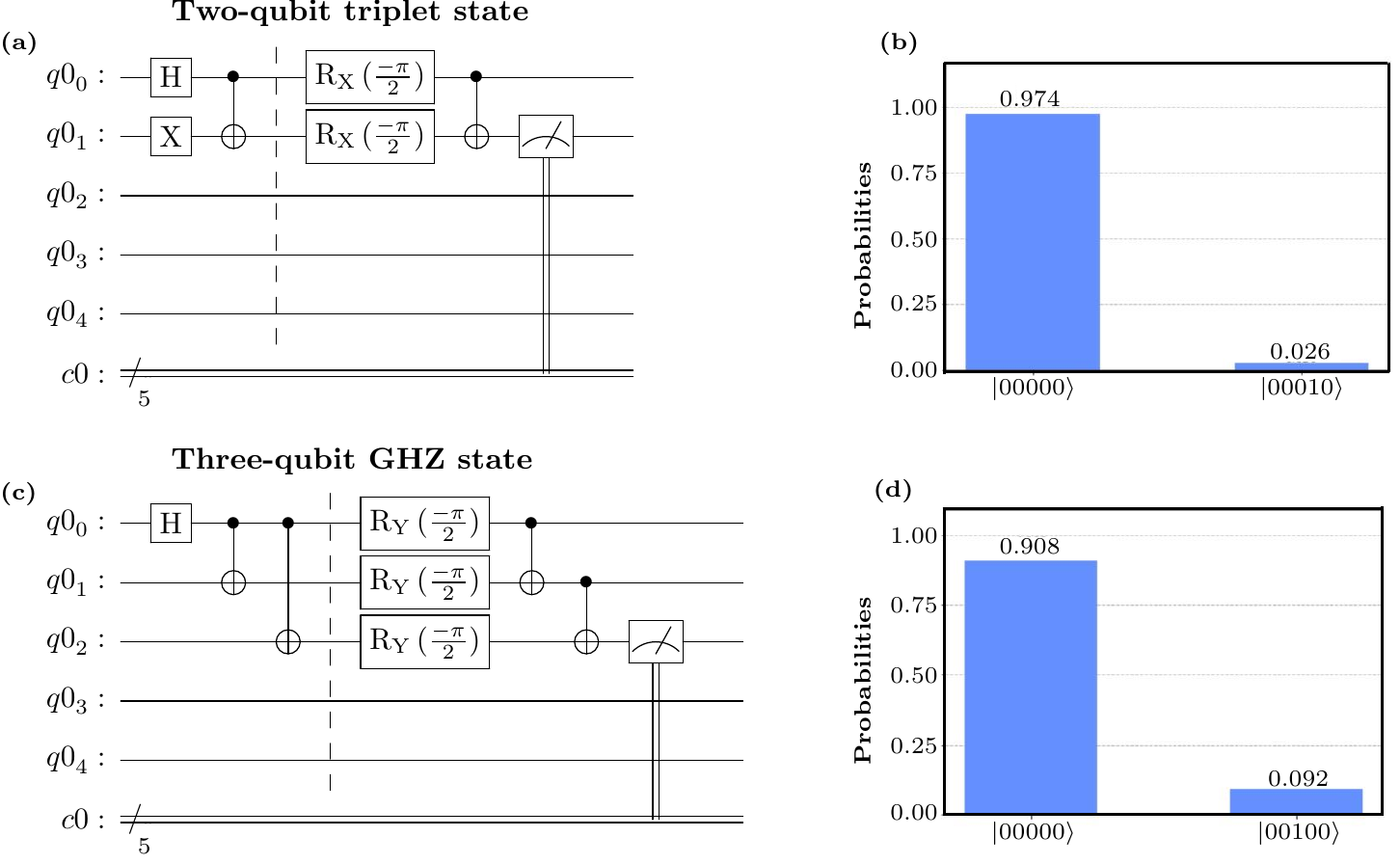}
\caption{IBM quantum circuit to create an entangled state. (a) The first block
of the circuit creates a two-qubit triplet state while the second block 
applies the quantum map $O_{2}$=CNOT.$\overline{X}_{2}$.$\overline{X}_{1}$ to
measure the 
expectation value of $\braket{\sigma_{1y}\sigma_{2y}}$ by detecting the
second qubit in the $z$ basis where
$\overline{X}_{1}=\overline{X}_{2}=R_{X}(\frac{-\pi}{2})$. (b) Histogram
representing probabilities of 
obtaining the second qubit in the $\ket{0}$ and the
$\ket{1}$ state with the values $p_{0}=0.974$ and $p_{1}=0.026$, respectively.
(c) The first block of the circuit creates a three-qubit GHZ state and
while the second block applies the quantum map
$P_{4}$=CNOT$_{23}$.$\overline{Y}_{3}$.CNOT$_{12}$.$\overline{Y}_{2}$.$\overline{Y}_{1}$
to measure the expectation value of 
$\braket{\sigma_{1x}\sigma_{2x}\sigma_{3x}}$ by
detecting the third qubit in the $z$ basis where
$\overline{Y}_{1}=\overline{Y}_{2}=\overline{Y}_{3}=R_{Y}(\frac{-\pi}{2})$. (d)
Histogram representing probabilities of obtaining the third qubit in the
$\ket{0}$ and the $\ket{1}$ state with the values $p_{0}=0.908$ and
$p_{1}=0.092$, respectively.}
\label{23q_cir} 
\end{figure*} 

The URDD sequence is designed for an even number of pulses where the propagator
of the $n$-pulse DD sequence is written as:
$U^{n}=U(\phi_{n})...U(\phi_{2})U(\phi_{1})$ where the phases
$\phi_{1}$,...,$\phi_{n}$ are the free control parameters that need 
to be 
appropriately designed 
and each pulse completely inverts the population i.e.  transition
probability $p=1$ in the ideal case~\citep{Genov-prl-2017}.  Thus, the
target propagator is defined as
$U_{0}=U^{(n)}(p=1)=(-1)^{n/2}\exp{(i\beta\widehat{I}_{z})}$ where
$\beta=2\sum_{k=1}^{n/2}(\phi_{2k}-\phi_{2k-1})$. The performance of the DD
sequence is measured by fidelity:

\begin{eqnarray}
F=\frac{1}{2}\vert Tr(U_{0}^{\dagger}U^{(n)})\vert=1-\delta_{n}
\end{eqnarray}
where $U_{0}$ is the target operator and $U^{(n)}$ is the operator of the DD
sequence that depends on relative phases between pulses $\phi_{k}$ (phase of
the $k$th pulse) and $\delta_{n}$ denotes the 
infidelity of $n$-pulse DD sequence.
The goal is to derive the control phases $\phi_{k}$ which can minimize the
infidelity of the DD sequence and 
for that the Taylor expansion is taken with respect
to errors in the control parameters by finding appropriate phases. 
The phases of the
$n$-pulse URDD sequence is given in the general form as \citep{Genov-prl-2017}:

\begin{equation}
\phi_{k}^{(n)}=\frac{(k-1)(k-2)}{2}\Phi^{(n)}+(k-1)\phi_{2}, 
\end{equation}
where
\begin{equation}
\Phi^{(4m)}=\pm \frac{\pi}{m},\quad \Phi^{(4m+2)}=\pm \frac{2m\pi}{2m+1}
\end{equation}
$k$ is the phase of $k$th pulse of the $n$-pulse URDD sequence 
and $\phi_{2}$ is
arbitrarily chosen.

\subsection{Standard DD Method}
\label{stu}
In the standard method, the DD sequence is repetitively applied as shown in
Fig.\ref{urdd8}(b) where each unit of the 
DD sequence corresponds to an identity operation in the
ideal condition;  however practically such operations are 
not possible due to the
presence of experimental errors. Thus, each unit of 
the DD sequence is written as~\citep{Zhen-prl-2019}:
\begin{equation}\label{un1}
U_{_{unit}}=
\begin{bmatrix}
1 & iC\epsilon\\
iC^{*}\epsilon & 1
\end{bmatrix}
+O(\epsilon^{2})
\end{equation}
where $C$ is a complex number that depends on the structure of
the DD sequence and
$\epsilon$ is the experimental error. When each DD unit is repeated M times as
$U_{_{M}}=U_{_{unit}}...U_{_{unit}}=(U_{_{unit}})^{M}$, the errors present in
one unit get accumulated, and the matrix form of the DD sequence
becomes~\cite{Zhen-prl-2019}:
\begin{equation}
\label{un2} 
U_{_{M}}= \begin{bmatrix} 1 & iMC\epsilon\\
iMC^{*}\epsilon & 1 \end{bmatrix} +O(\epsilon^{2}) 
\end{equation}
where
the off-diagonal terms of 
the matrix show that the errors get accumulated linearly upon
several applications of the 
DD sequence.

\subsection{Phase Randomization Method}
\label{prt} 
As explained in Sec.\ref{stu}, errors get accumulated
upon applying the basic unit of DD sequence $M$ times. In the phase
randomization (PR) technique \citep{Zhen-prl-2019}, a random global phase
$\Theta_{r,s}$ ($r$ denotes random phase and $s$ denotes $s$th unit of DD
sequence) is introduced to all $\pi$ pulses of $s$th unit of DD sequence and
the random phase changes each time the unit of DD sequence is repeated(
Fig.\ref{urdd8}(c)). The matrix form of the DD sequence
after adding the PR sequence is given by~\citep{Zhen-prl-2019}:
\begin{eqnarray}
\label{un3} 
U_{_{M}}& =
&U_{_{unit}}(\Theta_{r,M})...U_{_{unit}}
(\Theta_{r,2})U_{_{unit}}(\Theta_{r,1})\\
& = & \begin{bmatrix}
\label{un4} 1 & iZ_{r,M}MC\epsilon\\
iZ_{r,M}^{*}MC^{*}\epsilon & 1 \end{bmatrix} +O(\epsilon^{2}) 
\end{eqnarray}
where $Z_{r,M}=1/M\sum_{s=1}^{M}exp(-i\Theta_{r,s})$. The contribution of
$Z_{r,M}$ in the off-diagonal terms of the matrix 
serves to improve the robustness of the overall
DD sequence and the factor $Z_{r,M}$ act as a random walk 
due to the random phase
$\Theta_{r,s}$.  
\subsection{Correlated Phase Randomization Method}
\label{cprt} 
In the correlated phase randomization (CPR) technique
\citep{Zhenyu-symm-2020}, a condition is imposed in the PR technique where the
random phases should be chosen such that
$Z_{r,M}=1/M\sum_{s=1}^{M}exp(-i\Theta_{r,s})=0$. This condition erases the
off-diagonal terms of the matrix given in Eq.\ref{un4} and thus suppresses the
experimental errors more effectively as compared to PR technique.
The CPR technique is also efficient because it does not depend
on the value of $M$, and even smaller number of $M$s are sufficient to erase
the error, whereas the 
phase randomization technique requires a larger number of $M$s to
suppress the error.  
\section{Experimental
demonstration of protection of maximally entangled multipartite state on 
the IBM
quantum processor}
\label{exp} 
\subsection{Experimental details}
\label{ed} 
The
IBM quantum processor is a freely available online service that gives access
to superconducting transmon based qubits \citep{Devitt-pra-2016}. We used
the five-qubit $ibmq_{-}manila$ quantum processor, chosen 
as the average $T_{1}$ rate is greater than $T_{2}$
with phase damping being a dominant noise channel.
$T_{1}$ values for the qubits were between $126-151\mu s$ and
$T_{2}$ values were between $50-66\mu s$. Length of single qubit gate of
$ibmq_{-}manila$ quantum processor was $35.5ns$ while error of single qubit
gate was of the order $10^{-3}$ and readout errors were of the order
$10^{-2}$. 
Using  four qubits of the quantum processor, we
demonstrate the protection of two-qubit (triplet), three-qubit (GHZ-state) and
four-qubit (GHZ and cluster states) multipartite entangled states. 
The quantum
circuit to create an entangled state on the IBM processor are shown in
Fig.\ref{23q_cir}(a,c) and Fig.\ref{4q_cir}(a,c) where each
circuit was implemented 8192 times to compute the Born probabilities.

We
protect the entangled states using 
a URDD sequence of order 8, which we term the UR8DD
sequence. 
Mathematically, the total operator of one unit of the UR8DD sequence is
written as UR8DD $=F(\tau/2)U(\phi_{8})F(\tau)...F(\tau)U(\phi_{1})F(\tau/2)$
where $F$ corresponds to free evolution operation for time period $\tau$ and
$U(\phi_{k})$ denotes the $k$th $\pi$ pulse operation about the axis $\phi_{k}$.
The sequence of phases of $\pi$ pulses of UR8DD sequence are $\phi_{k}=(0,
\pi/2, 3\pi/2, \pi, \pi, 3\pi/2, \pi/2, 0)$. On 
the IBM quantum processor, 
the UR8DD pulses are applied using 
single-qubit quantum gates and an arbitrary single-qubit gate can be applied as
\begin{equation}\label{un2} G(\alpha,\beta,\gamma)= \begin{bmatrix}
\cos(\alpha/2) & -e^{i\gamma}\sin(\alpha/2)\\ e^{i\beta}\sin(\alpha/2) &
e^{i(\beta+\gamma)}\cos(\alpha/2) \end{bmatrix} \end{equation} where $k$th
$\pi$ pulse used in UR8DD sequence can be implemented as a single-qubit quantum
gate by specifying the parameters as
$U(\phi_{k})=G(\pi,\phi_{k}-\pi/2,\pi/2-\phi_{k})$. For the free evolution
between the pulses and to track the free evolution dynamics of entangled
states, identity gates are used. The total time period (T) for which one unit
of UR8DD sequence applied was 0.84$\mu s$  where free evolution delay time
between two pulses was 70$ns$ and total number of gates involved in one unit
was 24. To protect multipartite entangled states, UR8DD sequences were
implemented simultaneously on each qubit contributing in generating
entanglement of the state. One unit of the UR8DD sequence was 
applied several times
(maximum 9 times) to preserve the entanglement for 
a longer time. 

Further to
improve the performance of the phase randomization technique, 
we repeated the
experiments by adding random phases to all $\pi$ pulses of each unit of UR8DD
where the $k$th pulse of UR8DD sequence was applied along the axis
$\phi_{k}+\Theta_{r,m}$. The experiments were repeated a maximum of 9 times ($M=9$)
after adding phase randomization technique to UR8DD sequence.
Phase randomization technique can be improved further by
using correlated phase randomization technique where fully random phases are
replaced with correlated phases. This technique
suppresses the experimental errors more efficiently as compared to PR technique
and works well even for less number of repetition of UR8DD sequence. Thus, we
repeated the experiment by adding correlated phase to UR8DD sequence that
require minimum two times repetition of UR8DD sequence. In each experiment,
total number of correlated phases are divided into combinations of two sets of
phases and three sets of phases such that the condition
$Z_{r,M}=1/M\sum_{s=1}^{M}exp(-i\Theta_{r,s})=0$ is satisfied. For example,
if the UR8DD sequence is applied for 5 times, 
then in the first two units of the UR8DD sequence,
phases should be such that $Z_{r,2}=exp[-i(\Theta_{r,1}+\Theta_{r,2})]/2=0$
and similarly, in the next three units of the UR8DD sequence,
$Z_{r,3}=exp[-i(\Theta_{r,3}+\Theta_{r,4}+\Theta_{r,5})]/3=0$. Thus the
overall condition of CPR technique is satisfied. 

\subsection{Entanglement Witness}
\label{ew} 
An entanglement witness ($W$) can detect the presence of entanglement in a
quantum state by directly measuring the observables
\citep{horodecki-pla-1996,Otfried-pr-2009,guhne-prl-2006,zhang-pra-2007,Bourennane-prl-2004,sperling-prl-2013}.
A quantum state $\rho$ is an entangled state if there exists a Hermitian
operator ( called entanglement witness) such that $Tr\left[W \rho\right] <0$
and if $Tr\left[W \rho\right] \geq0$ then the quantum state is separable.
Therefore, the negative expectation of 
the witness operator $W$ signifies the presence
of entanglement. 

The presence of entanglement in quantum states at different
time points are measured by entanglement witness during free
evolution and while implementing the UR8DD sequence and after
adding phase and correlated randomization methods.
We measured the entanglement witness and plotted the bar graph of
entanglement parameter by taking negative of entanglement witness $(
\vartheta$=$-$Tr$\left[ W\rho\right])$.  

\begin{figure}[t]
\includegraphics[angle=0,scale=1]{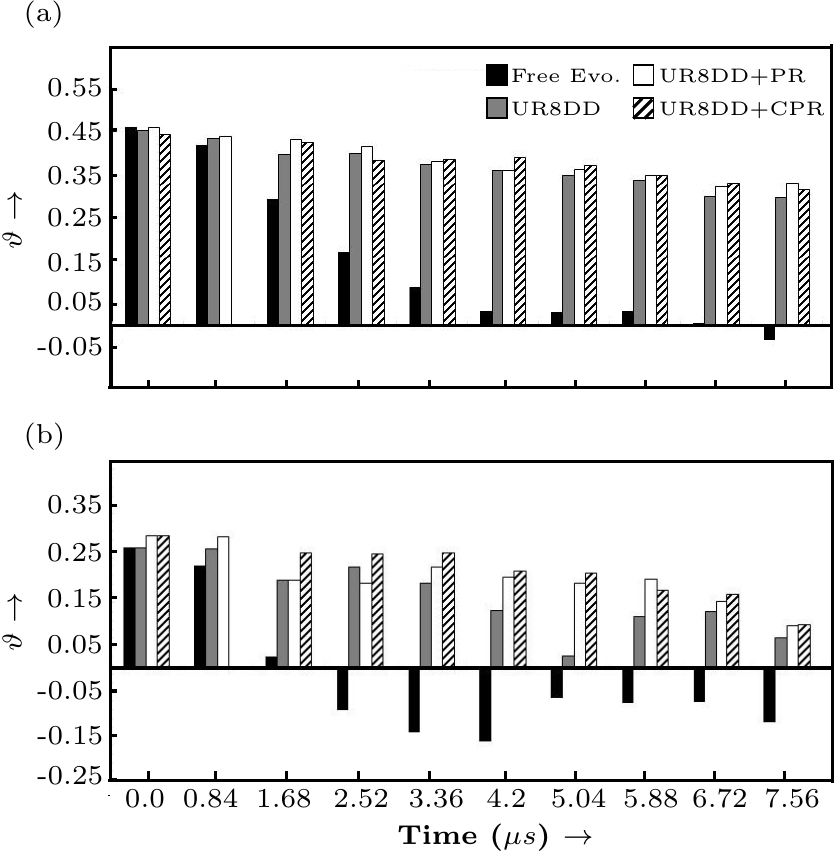} 
\caption{Bar plots of
the entanglement parameter ($\vartheta$) versus time ($\mu s$) of: (a) the
two-qubit triplet state and (b) the three-qubit GHZ state.  The black solid
bars represent $\vartheta$ of the states without applying any DD protection,
the grey solid bars represent $\vartheta$ of the states after applying the
UR8DD sequence, the white bars represent $\vartheta$ of the states after
applying the UR8DD sequence and adding the PR sequence (UR8DD+PR), and the
black cross-hatched bars represent $\vartheta$ of the states after applying the
UR8DD sequence and adding CPR sequence (UR8DD+CPR).} \label{23bar} \end{figure}
Thus, we experimentally measured 
the entanglement parameter of the two-qubit triplet
state, the three-qubit GHZ state and 
the four-qubit GHZ state and cluster states
at different time points, to quantify the preservation of 
entanglement in these states. 

\begin{figure*}[t]
\includegraphics[angle=0,scale=1]{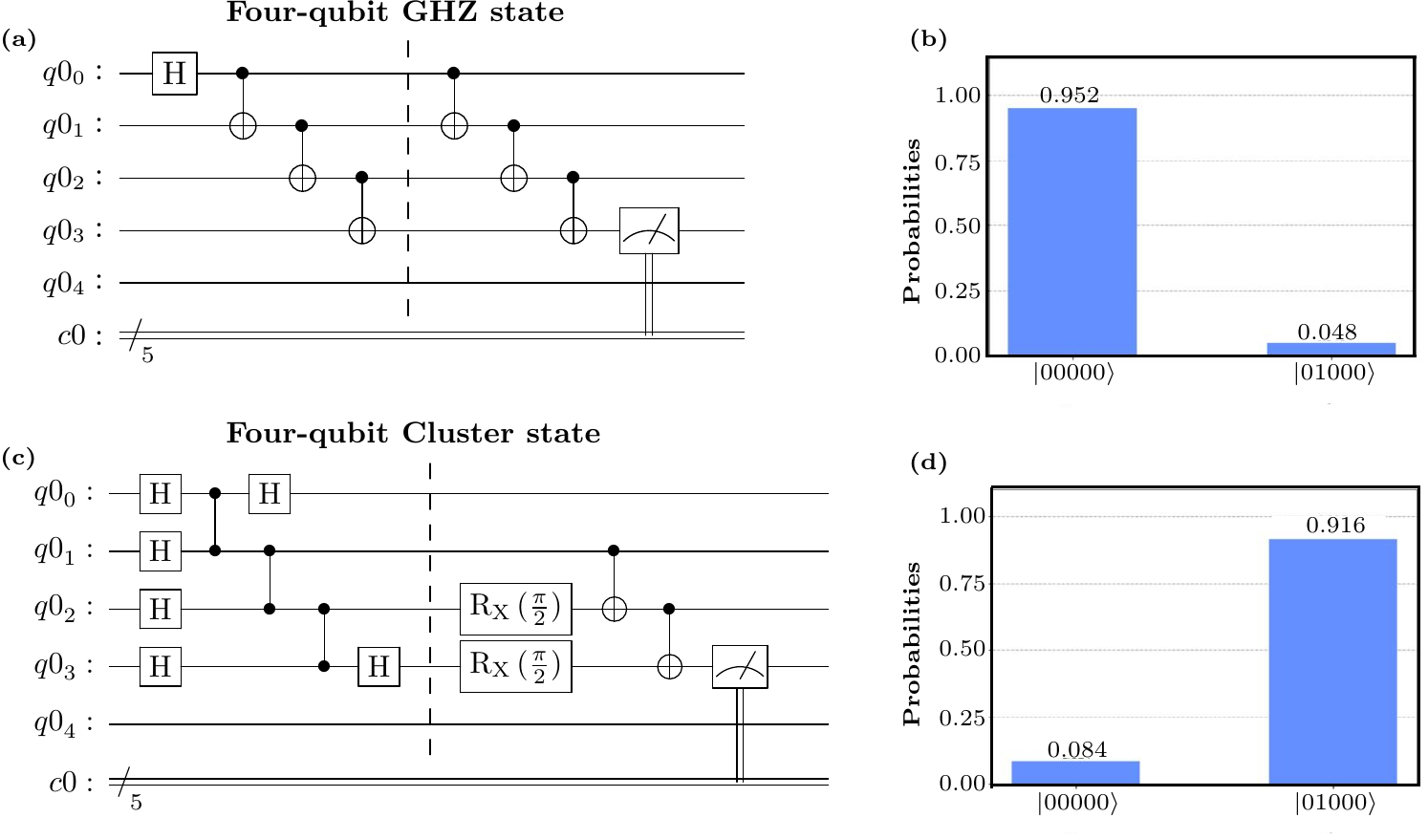}
\caption{The IBM quantum circuit to create four-qubit entangled states. (a) The
first block of the quantum circuit creates a four-qubit GHZ state and the
second block applies the quantum map
$Q_{7}$=CNOT$_{34}$.CNOT$_{23}$.CNOT$_{12}$ to measure the expectation value of
$\braket{\sigma_{1z}\sigma_{2z}\sigma_{3z}\sigma_{4z}}$ by detecting the fourth
qubit in the $z$ basis. (b) Histogram representing probabilities of obtaining
the fourth qubit in the $\ket{0}$ and the $\ket{1}$ state with the values
$p_{0}=0.952$ and $p_{1}=0.048$, respectively. (c) The first block of the
circuit creates a four-qubit cluster state and the second block applies the
quantum map $Q_{20}$=CNOT$_{34}$.$X_{4}$.CNOT$_{23}$.$X_{3}$ to measure the
expectation value of $\braket{\sigma_{2z}\sigma_{3y}\sigma_{4y}}$ by detecting
the fourth qubit in the $z$ basis where $X_{3}=X_{4}=R_{X}(\frac{\pi}{2})$. (d)
Histogram representing probabilities of obtaining the fourth qubit in the
$\ket{0}$ and the $\ket{1}$ state with the values $p_{0}=0.084$ and
$p_{1}=0.916$, respectively.} 
\label{4q_cir} 
\end{figure*}

\textbf{\textit{Two-qubit triplet state}}: We experimentally created 
the two-qubit
entangled state (triplet state) using two qubits  of five qubits
$ibmq_{-}manila$ quantum processor. The quantum circuit to create the state is
shown in first block of Fig.\ref{23q_cir}(a). We first studied the dynamics of
entanglement present in triplet state ($\rho_{TS}$) under free evolution and
then we protected the entanglement by implementing UR8DD sequence on both
qubits simultaneously. Further, we added phase randomization and correlated
phase randomization techniques to UR8DD sequence to enhance its performance in
extending the time period for which state can be protected. After each run of
DD sequence, the presence of entanglement is detected by measuring entanglement
witness directly where one run of DD sequence is of 0.84 $\mu s$. Entanglement
witness operator ($W_{TS}^{2}$) for triplet state can be written in linear
combination of $\sigma_{1x}\sigma_{2x}$, $\sigma_{1y}\sigma_{2y}$ and
$\sigma_{1z}\sigma_{2z}$ Pauli operator as \citep{Otfried-pr-2009}

\begin{equation}\label{2w}
W_{TS}^{2}=\frac{1}{4}(I_{2}-\sigma_{1x}\sigma_{2x}-\sigma_{1y}\sigma_{2y}+\sigma_{1z}\sigma_{2z})
\end{equation}
where $\lbrace\sigma_{x}$, $\sigma_{y}$, $\sigma_{z}\rbrace$ are single qubit Pauli basis. Experimentally, we measured witness operator $W_{TS}^{2}$ by measuring expectation value of three Pauli operators  where expectation of each Pauli operator was found experimentally by mapping the state $\rho$ to $\rho_{j}$ through unitary operator as $\rho_{j}=O_{j}\rho O_{j}^{\dagger}$ which is followed by observing $\braket{\sigma_{2z}}$ for the state $\rho_{j}$. Quantum circuit for one of the mapping is shown in second block of Fig.\ref{23q_cir}(a) and the details of mapping of Pauli operators to the single qubit $\sigma_{z}$ Pauli operator are given in Table \ref{2q_table} where $Y_{a}=R_{Y}(\frac{\pi}{2})$, $\overline{X}_{a}=R_{X}(\frac{-\pi}{2})$ and $a$ denotes the qubit number.
\begin{table} [h]
\caption{\label{2q_table}
Observables required to measure the entanglement witness for two qubits
(triplet state) are mapped to Pauli $\sigma_{z}$ operators through the initial
state transformation $\rho \rightarrow \rho_{j}=O_{j}\rho O_{j}^{\dagger}$.}
\begin{ruledtabular}
\begin{tabular}{cc}
\small{Observable expectations} & \small{Unitary operator}\\
\colrule
\scriptsize{$\braket{\sigma_{1x}\sigma_{2x}}$=Tr[$\rho_{1}$.$\sigma_{2z}$]} & \scriptsize{$O_{1}$=CNOT.Y$_{2}$.Y$_{1}$} \\
\scriptsize{$\braket{\sigma_{1y}\sigma_{2y}}$=Tr[$\rho_{2}$.$\sigma_{2z}$]} & \scriptsize{$O_{2}$=CNOT.$\overline{X}_{2}$.$\overline{X}_{1}$} \\
\scriptsize{$\braket{\sigma_{1y}\sigma_{2y}}$=Tr[$\rho_{3}$.$\sigma_{2z}$]} & \scriptsize{$O_{3}$=CNOT}
\end{tabular}
\end{ruledtabular}
\end{table}

Theoretically, the value of entanglement parameter $\vartheta$=$-$Tr$\left[
W_{TS}^{2}.\rho_{TS}\right] $ for two-qubit triplet state is $0.5$ and the
quantum state remains entangled until $\vartheta>0$. Experimentally, we found
the entanglement parameter of triplet state on IBM quantum processor as
$\vartheta_{exp}=0.461$. We plotted the bar graph of entanglement parameter
$\vartheta$ found experimentally at different time points corresponding to each
DD sequences (UR8DD, UR8DD+PR and UR8DD+CPR) and free evolution. The result is
shown in Fig.\ref{23bar}(a), from where it can be seen that UR8DD sequence has
protected the entanglement very well and its performance is improved further on
adding random phase and correlated random phase to it. Without
protection, entanglement dies at $7.56 \mu s$ and protection with UR8DD
sequence, $80\%$ of entanglement is preserved, with UR8DD+PR sequence ,
$72.1\%$ of entanglement is preserved whereas with UR8DD+CPR sequence, $71.1\%$
of entanglement is preserved.

\textbf{\textit{Three-qubit GHZ state}}: We experimentally constructed three-qubit GHZ state $(\rho_{GHZ}^{3})$ using three qubits of five qubits $ibmq_{-}manila$ quantum processor through the quantum circuit shown in first block of Fig.\ref{23q_cir}(c). We first studied the dynamics of entanglement present in three-qubit GHZ state under free evolution. Then we protected the quantum state by applying UR8DD sequence simultaneously on three qubits and its performance is improved by adding random phase and correlated random phase technique. 

\begin{table} [h]
\caption{\label{3q_table}
Observables required to measure the entanglement witness for three qubits (GHZ
state) are mapped to Pauli $\sigma_{z}$ operators through the initial state
transformation $\rho \rightarrow \rho_{j}=P_{j}\rho P_{j}^{\dagger}$.}
\begin{ruledtabular}
\begin{tabular}{cc}
\small{Observable expectations} & \small{Unitary operator}\\
\colrule
\scriptsize{$\braket{\sigma_{2z}\sigma_{3z}}$=Tr[$\rho_{1}$.$\sigma_{3z}$]} & \scriptsize{$P_{1}$=CNOT$_{23}$} \\
\scriptsize{$\braket{\sigma_{1z}\sigma_{3y}}$=Tr[$\rho_{2}$.$\sigma_{3z}$]} & \scriptsize{$P_{2}$=CNOT$_{13}$} \\
\scriptsize{$\braket{\sigma_{1z}\sigma_{2z}}$=Tr[$\rho_{3}$.$\sigma_{2z}$]} & \scriptsize{$P_{3}$=CNOT$_{12}$}\\
\scriptsize{$\braket{\sigma_{1x}\sigma_{2x}\sigma_{3x}}$=Tr[$\rho_{4}$.$\sigma_{3z}$]} & \scriptsize{$P_{4}$=CNOT$_{23}$.$\overline{Y}_{3}$.CNOT$_{12}$.$\overline{Y}_{2}$.$\overline{Y}_{1}$}\\
\scriptsize{$\braket{\sigma_{1x}\sigma_{2y}\sigma_{3y}}$=Tr[$\rho_{5}$.$\sigma_{3z}$]} & \scriptsize{$P_{5}$=CNOT$_{23}$.X$_{3}$.CNOT$_{12}$.X$_{2}$.$\overline{Y}_{1}$}\\
\scriptsize{$\braket{\sigma_{1y}\sigma_{2x}\sigma_{3y}}$=Tr[$\rho_{6}$.$\sigma_{3z}$]} & \scriptsize{$P_{6}$=CNOT$_{23}$.X$_{3}$.CNOT$_{12}$.$\overline{Y}_{2}$.X$_{1}$}\\
\scriptsize{$\braket{\sigma_{1y}\sigma_{2x}\sigma_{3y}}$=Tr[$\rho_{7}$.$\sigma_{3z}$]} & \scriptsize{$P_{7}$=CNOT$_{23}$.$\overline{Y}_{3}$.CNOT$_{12}$.X$_{2}$.X$_{1}$}\\
\end{tabular}
\end{ruledtabular}
\end{table}
To measure how well entanglement is protected, entanglement witness is measured experimentally at different time points without applying DD sequence and after applying DD sequence. Entanglement witness operator for three qubit GHZ state can be written in linear combination of seven Pauli operators as \citep{Otfried-pr-2009,guhne-ijtp-2003}
\begin{equation}\label{3w}
\begin{split}
W_{GHZ}^{3}=\frac{1}{8}(3I_{3}-\sigma_{2z}\sigma_{3z}-\sigma_{1z}\sigma_{3z}-\sigma_{1z}\sigma_{2z}-\sigma_{1x}\sigma_{2x}\sigma_{3x}\\+\sigma_{1x}\sigma_{2y}\sigma_{3y}+\sigma_{1y}\sigma_{2x}\sigma_{3y}+\sigma_{1y}\sigma_{2y}\sigma_{3x})
\end{split}
\end{equation}

Experimentally, expectation value of each Pauli operators was measured to
compute entanglement witness by mapping the state $\rho$ to $\rho_{j}$ through
unitary operator as $\rho_{j}=P_{j} \rho P_{j}^{\dagger}$ which was followed by
observing single qubit $\braket{\sigma_{2z}}$ or $\braket{\sigma_{3z}}$ for the
$\rho_{j}$. Quantum circuit for one of the mapping of three qubit is shown in
second block of Fig.\ref{23q_cir}(c) and the details of all Pauli operators
required to be measured for computing entanglement witness are given in Table
\ref{3q_table} where $X_{b}=R_{X}(\frac{\pi}{2})$,
$\overline{Y}_{b}=R_{Y}(\frac{-\pi}{2})$ and $b$ denotes the qubit number.
Theoretically, entanglement parameter is computed for three-qubit GHZ state
using entanglement witness as
$\vartheta$=$-$Tr$\left[W_{GHZ}^{3}\rho_{GHZ}^{3}\right]$ and its value is
$\vartheta=0.5$ while GHZ state remains entangled until $\vartheta>0$.
Experimentally, we got the entanglement parameter of initial three-qubit GHZ
state constructed on IBM quantum processor as $\vartheta_{exp}=0.298$. We
plotted the entanglement parameter $\vartheta$ versus time corresponding to
each DD sequence and free evolution as shown in Fig.\ref{23bar}(b). Our results
show that three-qubit GHZ state is very fragile as it can been seen that
entanglement parameter ($\vartheta$) decays  very fast under free evolution of
superconducting qubits and entanglement is protected very well on applying
UR8DD sequence and its performance is improved on adding PR and CPR technique.
Without protection, entanglement dies at $2.52 \mu s$ and on
protecting state with UR8DD sequence, $84.4\%$ of entanglement is preserved,
with UR8DD+PR sequence, $70.6\%$ of entanglement is preserved whereas with
UR8DD+CPR sequence, $94.8\%$ of entanglement is preserved.

\begin{figure}[t]
\includegraphics[angle=0,scale=1]{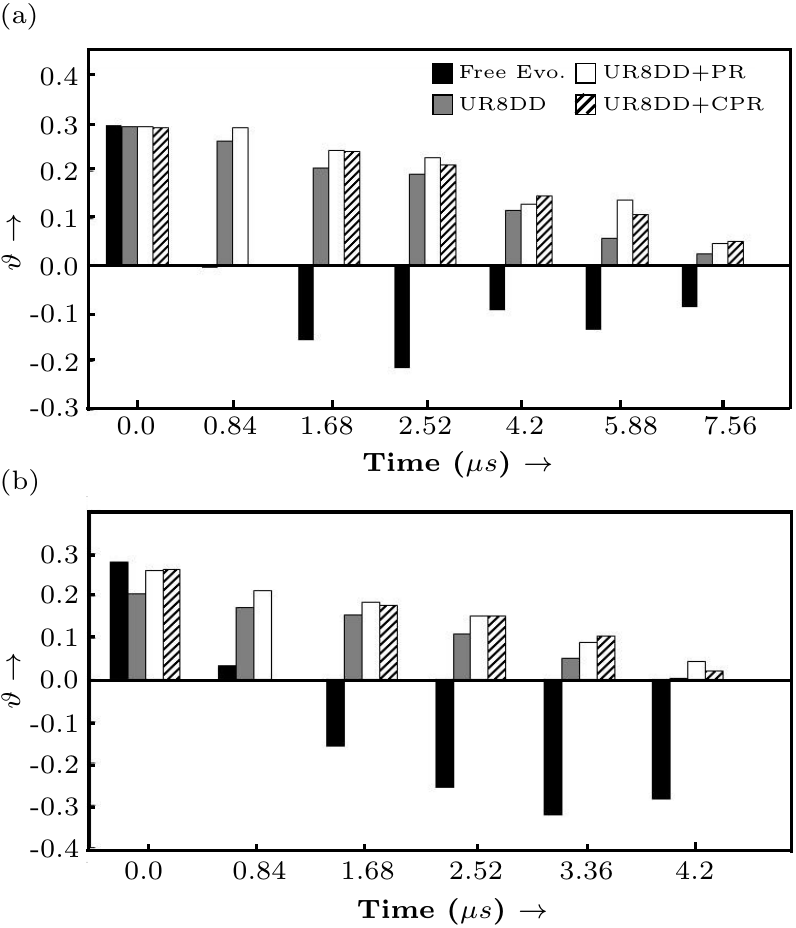}
\caption{Bar plots of the entanglement parameter ($\vartheta$) versus time
($\mu s$) of: (a) the four-qubit GHZ state and (b) the four-qubit cluster
state. The black solid bars represent $\vartheta$ of both states without
applying any DD protection, the grey solid bars represent $\vartheta$ of both
states after applying the UR8DD sequence, the white bars represent $\vartheta$
of both states  after applying the UR8DD sequence and adding a PR sequence
(UR8DD+PR) and the black cross-hatched bars represent $\vartheta$ of both
states  after applying the UR8DD sequence and adding a CPR sequence
(UR8DD+CPR).} 
\label{4bar} 
\end{figure}

\textbf{\textit{Four-qubit GHZ and cluster state}}: We next experimentally constructed four-qubit maximally entangled state, GHZ ($\rho_{GHZ}^{4}$) and Cluster state ($\rho_{CS}^{4}$) using four qubits of $ibmq_{-}manila$ quantum processor and quantum circuits to create GHZ and Cluster state are shown in Fig.\ref{4q_cir}(a) and Fig.\ref{4q_cir}(c) respectively. We studied the dynamics of free evolution of entanglement present in both four-qubit GHZ and Cluster state. We then protected both state by implementing UR8DD sequence and its variants on four qubits simultaneously. To observe how well entanglement is preserved in both states, entanglement witness was measured at different time points without applying DD sequence and after applying DD sequence. Entanglement witness operator for four-qubit GHZ state ($W_{GHZ}^{4}$) and Cluster state ($W_{CS}^{4}$) can be written as \citep{Bourennane-prl-2004,sperling-prl-2013,tokunaga-pra-2006}
\begin{equation}\label{4w1}
\begin{split}
W_{GHZ}^{4}=\frac{1}{16}(7I_{4}-E_{1}-E_{2}-E_{3}-E_{4}-E_{5}-E_{6}-E_{7}\\-E_{8}-E_{9}+E_{10}+E_{11}+E_{12}+E_{13}+E_{14}+E_{15})
\end{split}
\end{equation}  
\begin{equation}\label{4w2}
\begin{split}
W_{CS}^{4}=\frac{1}{16}(7I_{4}-E_{1}-E_{6}-E_{7}-E_{11}-E_{12}-E_{13}-E_{14}\\-E_{16}-E_{17}+E_{18}+E_{19}+E_{20}+E_{21}+E_{22}+E_{23})
\end{split}
\end{equation} 
where $E_{x}$'s are Pauli operator\\ 
$E_{1}=\sigma_{3z}\sigma_{4z}$,\quad$E_{9}=\sigma_{1y}\sigma_{2y}\sigma_{3y}\sigma_{4y}$,\quad$E_{17}=\sigma_{1z}\sigma_{3x}\sigma_{4x}$\\  
$E_{2}=\sigma_{2z}\sigma_{4z}$,\quad$E_{10}=\sigma_{1x}\sigma_{2x}\sigma_{3y}\sigma_{4y}$,\quad$E_{18}=\sigma_{1x}\sigma_{2x}\sigma_{4x}$\\  
$E_{3}=\sigma_{2z}\sigma_{3z}$,\quad$E_{11}=\sigma_{1x}\sigma_{2y}\sigma_{3x}\sigma_{4y}$,\quad$E_{19}=\sigma_{1x}\sigma_{2x}\sigma_{3z}$\\ 
$E_{4}=\sigma_{1z}\sigma_{4z}$,\quad$E_{12}=\sigma_{1x}\sigma_{2y}\sigma_{3y}\sigma_{4x}$,\quad$E_{20}=\sigma_{2z}\sigma_{3y}\sigma_{4y}$,\\
$E_{5}=\sigma_{1z}\sigma_{3z}$,\quad$E_{13}=\sigma_{1y}\sigma_{2x}\sigma_{3x}\sigma_{4y}$,\quad$E_{21}=\sigma_{1z}\sigma_{3y}\sigma_{4y}$,\\
$E_{6}=\sigma_{1z}\sigma_{2z}$,\quad$E_{14}=\sigma_{1y}\sigma_{2x}\sigma_{3y}\sigma_{4x}$,\quad$E_{22}=\sigma_{1y}\sigma_{2y}\sigma_{4z}$,\\   
$E_{7}=\sigma_{1z}\sigma_{2z}\sigma_{3z}\sigma_{4z}$,\quad$E_{15}=\sigma_{1y}\sigma_{2y}\sigma_{3x}\sigma_{4x}$,\\
$E_{8}=\sigma_{1x}\sigma_{2x}\sigma_{3x}\sigma_{4x}$,\quad$E_{16}=\sigma_{2z}\sigma_{3x}\sigma_{4x}$,\\
$E_{23}=\sigma_{1y}\sigma_{2y}\sigma_{3z}$\\
\\
Experimentally, expectation value of each Pauli  operator was  measured to compute entanglement witness of both state by mapping the state from $\rho$ to $\rho_{j}$ through unitary operator as $\rho_{j}=Q_{j}\rho Q_{j}^{\dagger}$ which was followed by observing single qubit $\braket{\sigma_{3z}}$ or $\braket{\sigma_{4z}}$ for $\rho_{j}$. Quantum circuit for one of the mapping of each four qubit GHZ state and Cluster state are shown in second block of Fig.\ref{4q_cir}(a) and Fig.\ref{4q_cir}(b) respectively and details of all Pauli operators are given in Table.\ref{4q_table}. Entanglement witness is used to compute entanglement parameter for both states and theoretically, which is found as $\vartheta$=$-$Tr$\left[W_{GHZ}^{4}\rho_{GHZ}^{4}\right]=0.5$ for GHZ state and $\vartheta$=-Tr$\left[W_{CS}^{4}\rho_{CS}^{4}\right]=0.5$ for Cluster State. Absence of entanglement is determined when entanglement parameter is $\vartheta\leq0$. Experimentally, entanglement parameter was computed for initial four-qubit GHZ and Cluster state as $\vartheta_{exp}=0.296$ and $\vartheta_{exp}=0.282$ respectively. 
\begin{table*} [t]
\caption{\label{4q_table}
Observables required to measure the entanglement witness for four qubits (GHZ
and cluster states) are mapped to Pauli $z$ operators through the initial state
transformation $\rho \rightarrow \rho_{j}=Q_{j}\rho Q_{j}^{\dagger}$.}
\begin{ruledtabular}
\begin{tabular}{cccc}
\small{Observable} & \small{Unitary operator} & \small{Observable} & \small{Unitary operator}\\
\colrule
\scriptsize{$E_{1}$=Tr[$\rho_{1}$.$\sigma_{4z}$]} & \scriptsize{$Q_{1}$=CNOT$_{34}$} & \scriptsize{$E_{13}$=Tr[$\rho_{13}$.$\sigma_{4z}$]} & \scriptsize{$Q_{13}$=CNOT$_{34}$.$X_{4}$.CNOT$_{23}$.$\overline{Y}_{3}$.CNOT$_{12}$.$\overline{Y}_{2}$.$X_{1}$}\\
\scriptsize{$E_{2}$=Tr[$\rho_{2}$.$\sigma_{4z}$]} & \scriptsize{$Q_{2}$=CNOT$_{13}$} & \scriptsize{$E_{14}$=Tr[$\rho_{14}$.$\sigma_{4z}$]} & \scriptsize{$Q_{14}$=CNOT$_{34}$.$\overline{Y}_{4}$.CNOT$_{23}$.$X_{3}$.CNOT$_{12}$.$\overline{Y}_{2}$.$X_{1}$}\\
\scriptsize{$E_{3}$=Tr[$\rho_{3}$.$\sigma_{3z}$]} & \scriptsize{$Q_{3}$=CNOT$_{23}$} & \scriptsize{$E_{15}$=Tr[$\rho_{15}$.$\sigma_{4z}$]} & \scriptsize{$Q_{15}$=CNOT$_{34}$.$\overline{Y}_{4}$.CNOT$_{23}$.$\overline{Y}_{3}$.CNOT$_{12}$.$X_{2}$.$X_{1}$}\\
\scriptsize{$E_{4}$=Tr[$\rho_{4}$.$\sigma_{4z}$]} & \scriptsize{$Q_{4}$=CNOT$_{14}$} & \scriptsize{$E_{16}$=Tr[$\rho_{16}$.$\sigma_{4z}$]} & \scriptsize{$Q_{16}$=CNOT$_{34}$.$\overline{Y}_{4}$.CNOT$_{23}$.$\overline{Y}_{3}$}\\
\scriptsize{$E_{5}$=Tr[$\rho_{5}$.$\sigma_{3z}$]} & \scriptsize{$Q_{5}$=CNOT$_{13}$} & \scriptsize{$E_{17}$=Tr[$\rho_{17}$.$\sigma_{4z}$]} & \scriptsize{$Q_{17}$=CNOT$_{34}$.$\overline{Y}_{4}$.CNOT$_{13}$.$\overline{Y}_{3}$}\\
\scriptsize{$E_{6}$=Tr[$\rho_{6}$.$\sigma_{2z}$]} & \scriptsize{$Q_{6}$=CNOT$_{12}$} & \scriptsize{$E_{18}$=Tr[$\rho_{19}$.$\sigma_{3z}$]} & \scriptsize{$Q_{19}$=CNOT$_{23}$.CNOT$_{12}$.$\overline{Y}_{2}$.$\overline{Y}_{1}$}\\
\scriptsize{$E_{7}$=Tr[$\rho_{7}$.$\sigma_{4z}$]} & \scriptsize{$Q_{7}$=CNOT$_{34}$.CNOT$_{23}$.CNOT$_{12}$} & \scriptsize{$E_{19}$=Tr[$\rho_{18}$.$\sigma_{4z}$]} & \scriptsize{$Q_{18}$=CNOT$_{24}$.CNOT$_{12}$.$\overline{Y}_{2}$.$\overline{Y}_{1}$}\\
\scriptsize{$E_{8}$=Tr[$\rho_{8}$.$\sigma_{4z}$]} & \scriptsize{$Q_{8}$=CNOT$_{34}$.$\overline{Y}_{4}$.CNOT$_{23}$.$\overline{Y}_{3}$.CNOT$_{12}$.$\overline{Y}_{2}$.$\overline{Y}_{1}$} & \scriptsize{$E_{20}$=Tr[$\rho_{20}$.$\sigma_{4z}$]} & \scriptsize{$Q_{20}$=CNOT$_{34}$.$X_{4}$.CNOT$_{23}$.$X_{3}$}\\
\scriptsize{$E_{9}$=Tr[$\rho_{9}$.$\sigma_{4z}$]} & \scriptsize{$Q_{9}$=CNOT$_{34}$.$X_{4}$.CNOT$_{23}$.$X_{3}$.CNOT$_{12}$.$X_{2}$.$X_{1}$} & \scriptsize{$E_{21}$=Tr[$\rho_{21}$.$\sigma_{4z}$]} & \scriptsize{$Q_{21}$=CNOT$_{34}$.$X_{4}$.CNOT$_{13}$.$X_{3}$}\\
\scriptsize{$E_{10}$=Tr[$\rho_{10}$.$\sigma_{4z}$]} & \scriptsize{$Q_{10}$=CNOT$_{34}$.$X_{4}$.CNOT$_{23}$.$X_{3}$.CNOT$_{12}$.$\overline{Y}_{2}$.$\overline{Y}_{1}$} & \scriptsize{$E_{22}$=Tr[$\rho_{22}$.$\sigma_{4z}$]} & \scriptsize{$Q_{22}$=CNOT$_{24}$.CNOT$_{12}$.$X_{2}$.$X_{1}$}\\
\scriptsize{$E_{11}$=Tr[$\rho_{11}$.$\sigma_{4z}$]} & \scriptsize{$Q_{11}$=CNOT$_{34}$.$X_{4}$.CNOT$_{23}$.$\overline{Y}_{3}$.CNOT$_{12}$.$X_{2}$.$\overline{Y}_{1}$} & \scriptsize{$E_{23}$=Tr[$\rho_{23}$.$\sigma_{3z}$]} & \scriptsize{$Q_{23}$=CNOT$_{23}$.CNOT$_{12}$.$X_{2}$.$X_{1}$}\\
\scriptsize{$E_{12}$=Tr[$\rho_{12}$.$\sigma_{4z}$]} & \scriptsize{$Q_{12}$=CNOT$_{34}$.$\overline{Y}_{4}$.CNOT$_{23}$.$X_{3}$.CNOT$_{12}$.$X_{2}$.$\overline{Y}_{1}$}\\
\end{tabular}
\end{ruledtabular}
\end{table*}

We plotted the bar graph of the entanglement parameter versus time for the GHZ
state and the cluster state corresponding to each DD sequence (UR8DD, UR8DD+PR,
UR8DD+CPR) and for free evolution without protection (shown in
Fig.\ref{4bar}(a) and Fig.\ref{4bar}(b)). From the bar graph, we observe that
the four-qubit GHZ and cluster states are very fragile as compared to the
two-qubit triplet state and the three-qubit GHZ state and hence the
entanglement parameter of these states decays very quickly. On applying the
UR8DD sequence, the entanglement is protected very well and the preservation is
improved upon applying variants of the UR8DD sequence, namely, the PR and CPR
techniques.  For the four-qubit GHZ state, the entanglement dies at $1.68 \mu
s$ and on protecting the state with UR8DD sequence, $70\%$ of entanglement is
preserved, while with the UR8DD+PR sequence, $82.1\%$ of the entanglement is
preserved, whereas the UR8DD+CPR sequence is able to preserve $81.4\%$ of the
entanglement. Similarly, for the four-qubit cluster state, the entanglement
dies at $1.68 \mu s$, and on protecting the state with UR8DD sequence, $54.8\%$
of entanglement is preserved, while with UR8DD+PR sequence, $65.5\%$ of the
entanglement is preserved, whereas  the UR8DD+CPR sequence is able to preserve
$62.6\%$ of the entanglement.

\section{Conclusions}
\label{con}
We have experimentally demonstrated the efficicay of the URDD sequence in
preserving multipartite entangled states generated on the IBM quantum computer
using superconducting qubits. Our results show that the UR8DD sequence is able
to successfully protect the entanglement present in such entangled states. The
entanglement witness was utilized to quantify the entanglement present in the
state at several time points, and to directly measure the entanglement present
via measurement on only a single qubit.  Further,  we show that the performance
of the basis UR8DD sequence can be substantially improved by adding the PR and
CPR variants.

Our work is a step forward in demonstrating the utility of URDD sequences in
protecting multipartite entanglement on NISQ computers. Future applications of
this work include integrating the URDD sequence with multiqubit quantum gates
during free evolution, for gate optimization and
protection~\citep{souza-pra-2012,Zhang-prl-2014}. 

\begin{acknowledgments} 
Authors acknowledge the support from IBM quantum experience team for providing
the experimental platform.  Arvind acknowledges financial support from
DST/ICPS/QuST/Theme-1/2019/General Project number Q-68.  K. D. acknowledges
financial support from DST/ICPS/QuST/Theme-2/2019/General Project number Q-74.
\end{acknowledgments}


%

\end{document}